\documentclass[authoryear,preprint,review,12pt]{elsarticle}

\usepackage{amssymb}

\usepackage{lineno}

\journal{Icarus}

\begin{document}

\begin{frontmatter}


 \title{Fragment properties from large-scale asteroid collisions: I: Results from SPH/N-body simulations using porous parent bodies and improved material models}
 

\author[label1]{Martin Jutzi}
\address[label1]{Physics Institute, University of Bern, NCCR PlanetS, Gesellsschaftsstrasse 6, 3012 Bern, Switzerland}
\ead{martin.jutzi@space.unibe.ch}

\author[label2]{Patrick Michel}
\address[label2]{Universite C\^ote dÕAzur, Observatoire de la C\^ote dÕAzur, Centre National de la Recherche Scientifique, Laboratoire Lagrange, Nice, France. }

\author[label3]{Derek C. Richardson}
\address[label3]{Department of Astronomy, University of Maryland, College Park, MD, USA}




\begin{abstract}
Understanding the collisional fragmentation and subsequent reaccumulation of fragments is crucial for studies of the formation and evolution of the small-body populations. 
Using an SPH / N-body approach, we investigate the size-frequency distributions (SFDs) resulting from the disruption of 100 km-diameter targets consisting of porous material, including the effects of pore-crushing as well as friction. Overall, the porous targets have a significantly higher impact strength ($Q^*_D$) than the rubble-pile parent bodies investigated previously \citep{Benavidez:2012ez} and show a behavior more similar to non-porous monolithic targets \citep{Durda:2007db}.
Our results also confirm that for a given specific impact energy, the SFDs resulting from a parent body disruption are strongly dependent on the size scale.

\end{abstract}


\end{frontmatter}


\newpage

\section{Introduction}\label{s:intro}

The observed asteroid families are composed of bodies that are thought to have originated from energetic collisions, which in turn lead to disruptions of larger parent bodies \citep[e.g.][]{Farinella:1996fd}. Understanding the collisional fragmentation and the subsequent reaccumulation of fragments is crucial for studies of processes taking place during the formation of the solar system and to reconstruct the internal structure of small bodies.
As a complement to experimental and theoretical approaches, numerical modeling has become an important component to study asteroid collisions and impact processes \citep[e.g.][]{Jutzi:2015ux,Michel:2015mr}. 
The process of large-scale disruptions consists of two distinct phases: the impact and fragmentation phase, and the gravitational reaccumulation phase. They are characterized by very different time scales and therefore can be studied by a hybrid modeling approach, coupling shock-physics code models and gravitational N-body methods \citep[e.g.][]{Michel:2001mb,Michel:2003mb}.
In numerous modeling studies, the effects of various target properties and impact conditions on the outcome of disruptive asteroid collisions have been investigated. Monolithic, pre-shattered, micro-porous or rubble-pile targets \citep[e.g.][]{Michel:2003mb,Durda:2007db,Jutzi:2010bf,Benavidez:2012ez} have been studied and different parent body sizes have been explored \citep[e.g.][]{Benavidez:2018bd,Sevecek:2017sb}. Effects related to the numerical scheme (such as the resolution) have been investigated as well \citep{Genda:2015gf,Genda:2017gf}. Recently, the collisional disruption of planetesimals in the gravity regime has also been explored with the grid-based iSALE code \citep{Suetsugu:2018st}. 

The comparison between simulation outcomes for various kinds of parent-body structures and the observed properties of asteroid family properties can help to constrain the internal structures of the parent body of the considered families.

The size-frequency distributions (SFDs) resulting from the disruption of 100 km-diameter targets have been determined for bodies consisting of either monolithic non-porous basalt \citep{Durda:2007db} or non-porous basalt blocks held together by gravity ('rubble piles') \citep{Benavidez:2012ez}. Here we use the same  range of collision speeds, impact angles, and specific impact energies and extend those studies to targets consisting of micro-porous material. Recent studies have shown that the presence of microporosity influences the outcome of a catastrophic disruption \citep{Jutzi:2008kp,Jutzi:2010bf}. Many asteroid families are of dark taxonomic type, such as C-type, which is often considered to contain a high fraction of porosity (including microporosity) based on measured bulk densities of C-type asteroids \citep[e.g.][]{Britt:2002wx}. Therefore, to determine the impact conditions for the formation of dark-type asteroid families, a comparison is needed between the actual family SFDs and those of impact disruptions of porous bodies. Moreover, the comparison between the disruptions of non-porous, rubble-pile, and porous targets is important to assess the influence of various internal structures on the outcome.

For the modeling of the impact and fragmentation phase we use a shock-physics code based on the Smoothed  Particle Hydrodynamics (SPH) technique. The code includes a well-tested porosity model as well as updated strength and friction models (section \ref{s:shockphysicscode}), which were not included in the previous studies  \citep{Durda:2007db,Benavidez:2018bd,Sevecek:2017sb}. It has been shown recently that the effects of porosity as well as friction can lead to significant differences in the outcome of asteroid collisions \citep{Jutzi:2015gb}. In addition to the effect of material properties, we also investigate the dependence of the SFDs on the parent body size. 

In section \ref{s:methods}, our model approach is described and the differences compared to previous studies are indicated. The results of our study are presented in section \ref{s:results}; conclusions and outlook are in section \ref{s:conclusions}.
	
\section{Model approach}\label{s:methods}
In this section we describe our modeling approach as well as the assumptions regarding the internal structures and initial conditions. 

\subsection{Numerical method}
To model the collision process and subsequent reaccumulation, we use an SPH / N-body approach as introduced by \citet{Michel:2001mb,Michel:2003mb}. This modelling approach has been applied in a number of recent studies  \citep[e.g.][]{Durda:2007db,Benavidez:2012ez,Benavidez:2018bd,Sevecek:2017sb} using the original method. However, both the SPH shock-physics codes as well as the N-body code pkdgrav have been extended and improved significantly in recent years. Here we briefly describe the basic methods and recent improvements. 

\subsubsection{Shock-physics code}\label{s:shockphysicscode}
We use a parallel (distributed memory) SPH impact code \citep{Benz:1994ij,Benz:1995hx,Nyffeler:2004tz,Jutzi:2008kp,Jutzi:2015gb} that includes self-gravity as well as material strength models. To model fractured, granular material, a pressure-dependent shear strength (friction) is included by using a standard Drucker-Prager yield criterion \citep{Jutzi:2015gb}. In most previous SPH / N-body simulations, fully damaged material was treated as a strengthless ÔfluidÕ, which can lead to a significant underestimation of the 'impact strength' of the target asteroid \citep{Jutzi:2015gb}. The effect of friction can also lead to increased impact heating \citep{Kurosawa:2018kg}. Porosity is modeled using a sub-resolution approach based on the P-alpha model  \citep{Jutzi:2008kp}. The material properties  (crush-curve) of the porous target used here are those that provided the best match to impact experiments on pumice targets \citep{Jutzi:2009ht}. The porosity model takes into account the enhanced dissipation of energy during compaction of porous materials, an effect not included in the 'rubble-pile' models used by \citet{Benavidez:2012ez} and \citet{Benavidez:2018bd}. We further use the Tillotson Equation of State (EOS) with parameters for basalt (except for the density) as given in \citet{Jutzi:2009ht}. 

\subsubsection{N-body code pkdgrav}
We use the same procedure as in previous papers \citep{Michel:2003mb,Jutzi:2010bf}. Fragments represented by SPH particles in the previous phase are replaced by spherical particles that can interact under their mutual gravity, collide and bounce or merge when their relative speed is larger or smaller than their mutual escape speed. In case of merger, the two particles are replaced by a spherical particle with the same momentum. In case of bouncing, a normal coefficient of restitution set to $0.3$ and a tangential coefficient of restitution set to $1$ are used to model an inelastic collision between two porous fragments (hence the choice of a rather low normal coefficient; see also \citet{Jutzi:2010bf}). This approach prevents us from obtaining information on the shape of reaccumulated fragments, but allows us to obtain their size and ejection velocity distributions, which is our main interest in this paper, for comparison with previous studies.

\subsubsection{Handoff between the two methods}
Once the fragmentation phase is over, the hydrodynamical simulations are stopped and the SPH particles and their corresponding velocity distribution are fed into the N-body code that computes the dynamical evolution of the system to late time. For the transfer time we use $t$ = 400 s, except for the case of the large 200 km target (section \ref{sec:inicond}) where $t$ = 1200 s is used. This procedure is the same as used previously \citep[e.g.][]{Jutzi:2010bf}. The gravitational phase was carried out to a simulated time of about 12 days, after which the outcome essentially does not change anymore.

\subsection{Model of the internal structure}
As noted in section \ref{s:intro}, we consider porous parent bodies in this study. The scale of porosity is defined in comparison with the other relevant dimensions involved in the problem, such as the size of the projectile and/or crater. Using a sub-resolution P-alpha type porosity model implicitly assumes that the scale of the porosity is smaller than the scale of the impactors (i.e., the scale of porosity is assumed to be smaller than a few 100 meters). 

A body containing such small-scale porosity may be crushable: cratering on a microporous asteroid is an event involving compaction rather than ejection \citep{Housen:1999iz}. Thus, for an impact into a microporous material, a part of the kinetic energy is dissipated by compaction, which leads to less ejected mass and lower speeds of the ejected material. These effects cannot be reproduced by hydrocodes developed for the modeling of non-porous solids.

In contrast, \citet{Benavidez:2012ez,Benavidez:2018bd} used targets that were constructed by filling the interior of a spherical shell with an uneven distribution of non-porous basalt spheres, leading to a structure with large-scale voids. 

\subsection{Initial conditions}\label{sec:inicond}
We use the same matrix of impact conditions as explored in \citet{Durda:2007db}  and \citet{Benavidez:2012ez}, covering a wide range of impact speeds (from 3 to 7 km/s), impact angles (from 15$^\circ$ to 75$^\circ$ with 15$^\circ$ increments) and impactor diameters (chosen to obtain the same range of specific impact energies as in the previous studies). We use non-porous basalt impactors with initial densities of 2.7 g/cm$^3$ and porous targets with an initial density of 1.3 g/cm$^3$, corresponding to a porosity of $\sim$ 50\%.  

In addition, we perform a few exploratory runs at different scales (target diameters ranging from 25 km to 200 km, with scaled impactor sizes). For our nominal simulations, a resolution of 4$\times$10$^5$ particles is used. As recently shown by \citet{Genda:2015gf}, the outcome of disruption simulations (i.e. the catastrophic disruption threshold $Q^*_D$) depends on the numerical resolution, in particular for very low particle numbers. With the moderately high resolution used here, the outcomes are reasonably close to convergence, given the much larger effects of material properties (strength/friction, porosity) on the outcome \citep[e.g.][]{Jutzi:2015gb}. Although a deeper investigation may be needed, a few runs performed with 1$\times$10$^5$ and 1$\times$10$^6$ particles showed similar mass ratios of the largest remnant to the parent body to that obtained with 4$\times$10$^5$ particles.

\section{Results}\label{s:results}
\subsection{Size of largest remnant}
\begin{figure}
\begin{center}
\includegraphics[width=6.5cm]{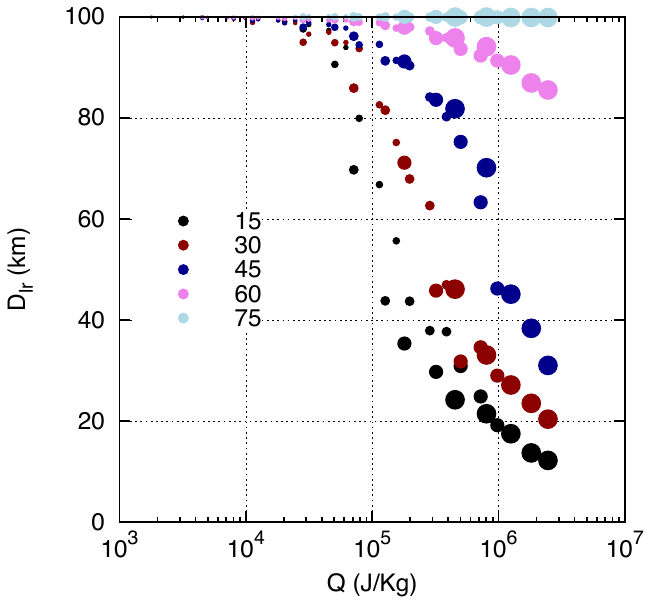}\\
\includegraphics[width=6cm]{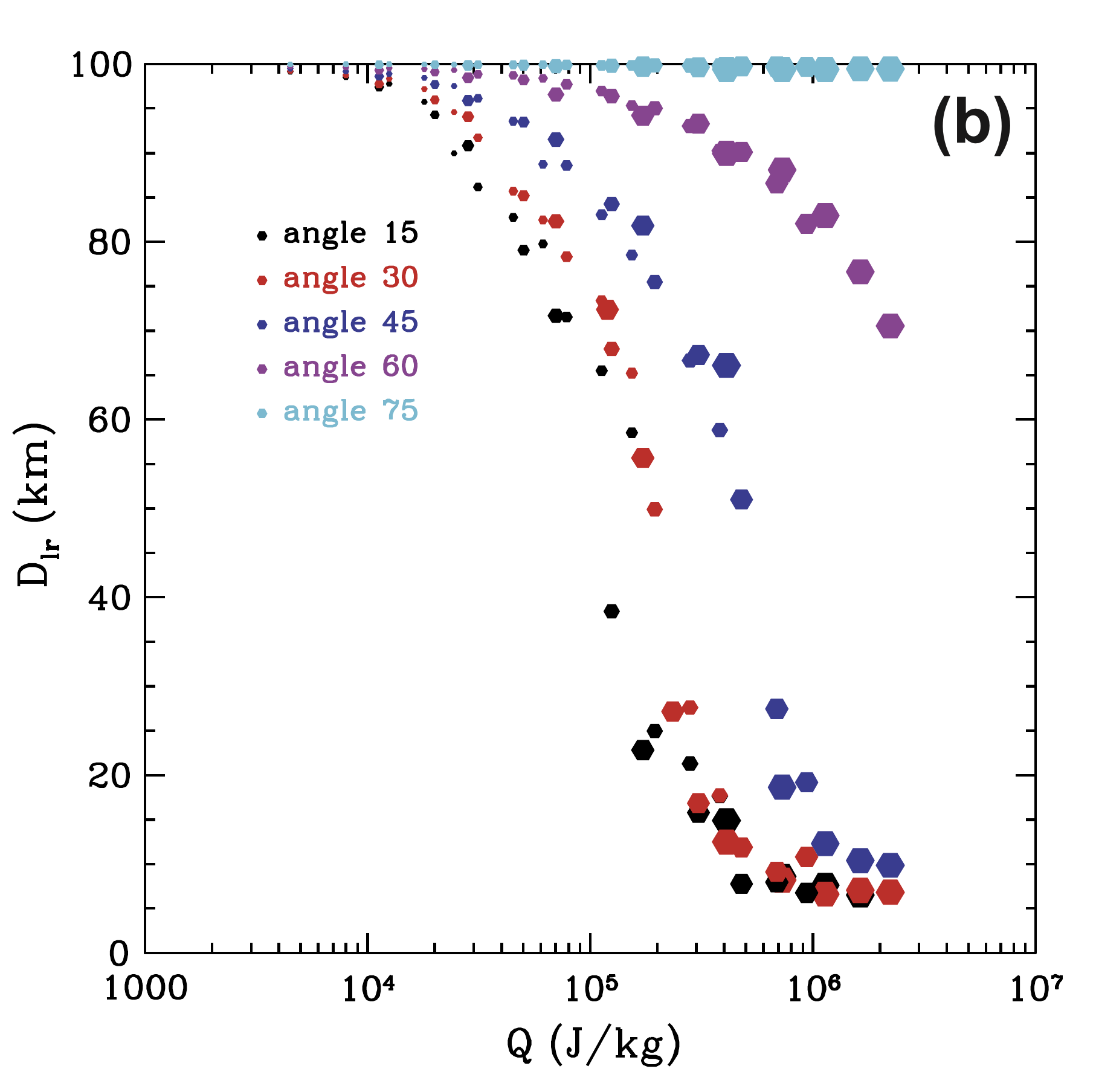}
\includegraphics[width=6cm]{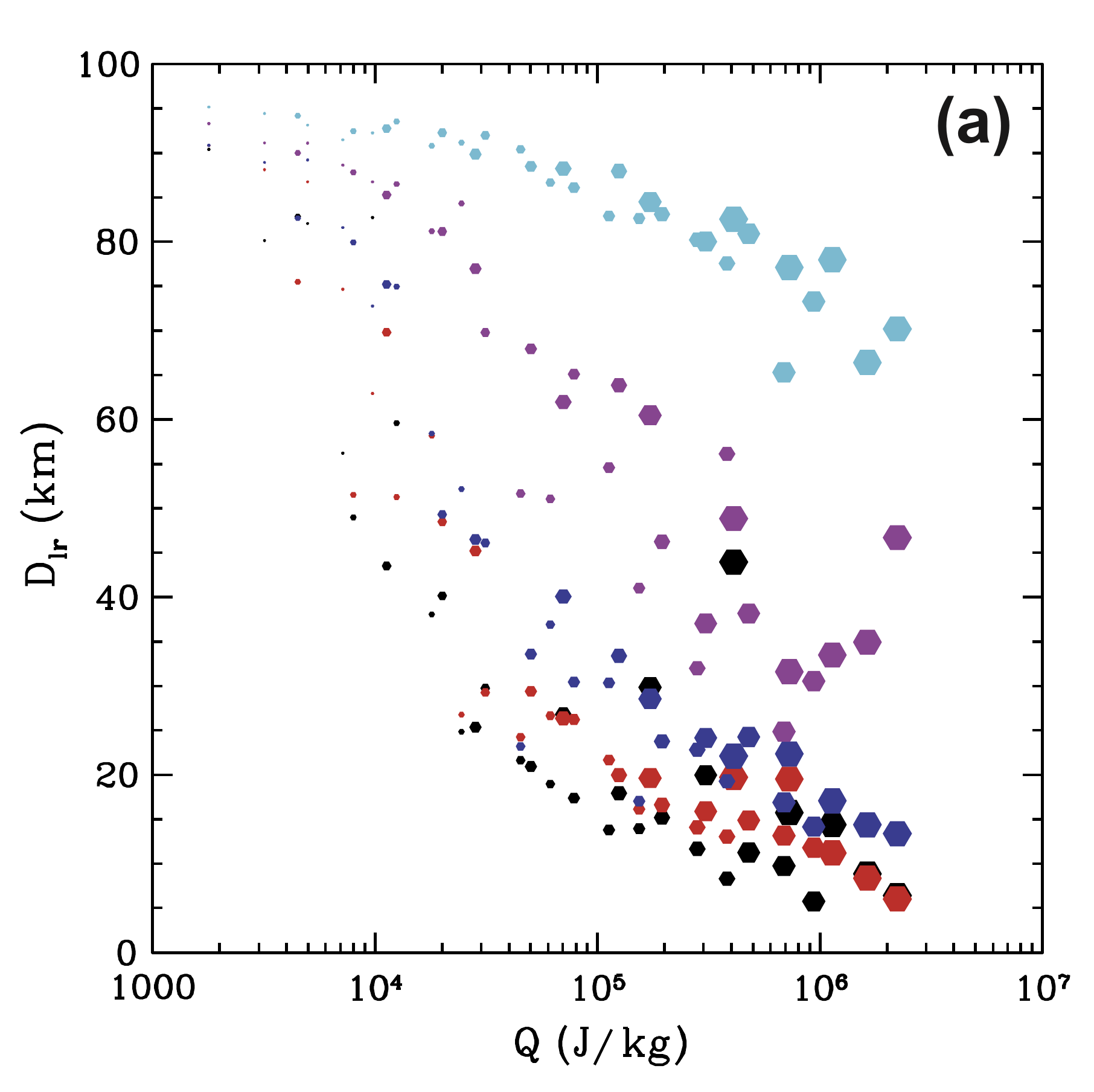}
\caption{Size of largest fragment as a function of specific impact energy. Note: projectile material is not included in the mass computations. Top: results of this study; bottom left: monolithic targets \citep{Durda:2007db}; bottom right: rubble-pile targets \citep{Benavidez:2012ez}. We note that for a given specific impact energy, the projectile sizes are slightly different from the previous studies due to different target densities. }
\label{fig:mlrq}
\end{center}
\end{figure}

An overview of the simulation results in terms of the size of the largest remnant for each collision is shown in Figure \ref{fig:mlrq} as a function of the specific impact energy. The results are compared to the results obtained in previous studies using non-porous monolithic \citep{Durda:2007db}  and rubble-pile \citep{Benavidez:2012ez} parent bodies. Overall, the porous targets investigated here show a behavior more similar to the non-porous monolithic targets than the rubble-pile ones. It has already been suggested in \citet{Jutzi:2010bf} that, in the gravity regime, there is not a large difference between $Q^*_D$ for porous and non-porous materials because different effects compensate for each other (e.g. energy dissipation by compaction for porous targets vs. higher material strength and stronger gravity because of higher density in non-porous targets). On the other hand, our results show that the porous targets considered here have a significantly higher impact strength than the rubble-pile targets used by \citet{Benavidez:2012ez}, confirming the findings by \citet{Jutzi:2015gb}. As argued by \citet{Jutzi:2015gb}, the rubble-pile targets as modeled by \citet{Benavidez:2012ez} behave more like porous fluids rather than real rubble-pile bodies. This is because friction of fully damaged material is not included in their SPH model, which therefore may have omitted an important effect governing granular flow.\\

\subsection{Size distributions}
\subsubsection{Results for $D$ = 100 km parent body}
\begin{figure}
\begin{center}
\includegraphics[width=14cm]{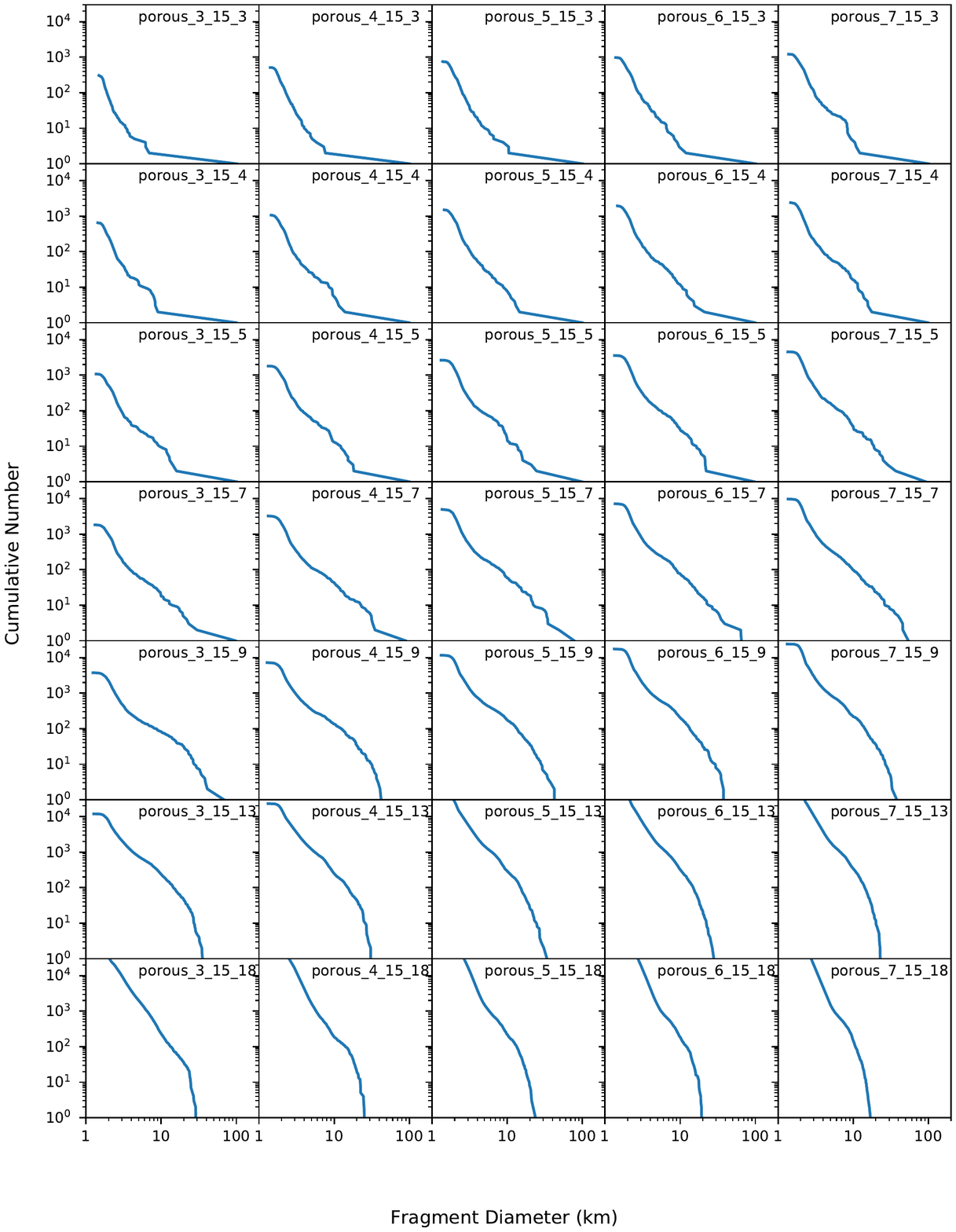}
\caption{Cumulative size distributions for impact angle $\theta$ = 15$^{\circ}$.  For each run, the first number of the legend is the impact velocity in km/s, the second the impact angle and the third the approximate projectile radius in km.}
\label{fig:sfd15}
\end{center}
\end{figure}

\begin{figure}
\begin{center}
\includegraphics[width=14cm]{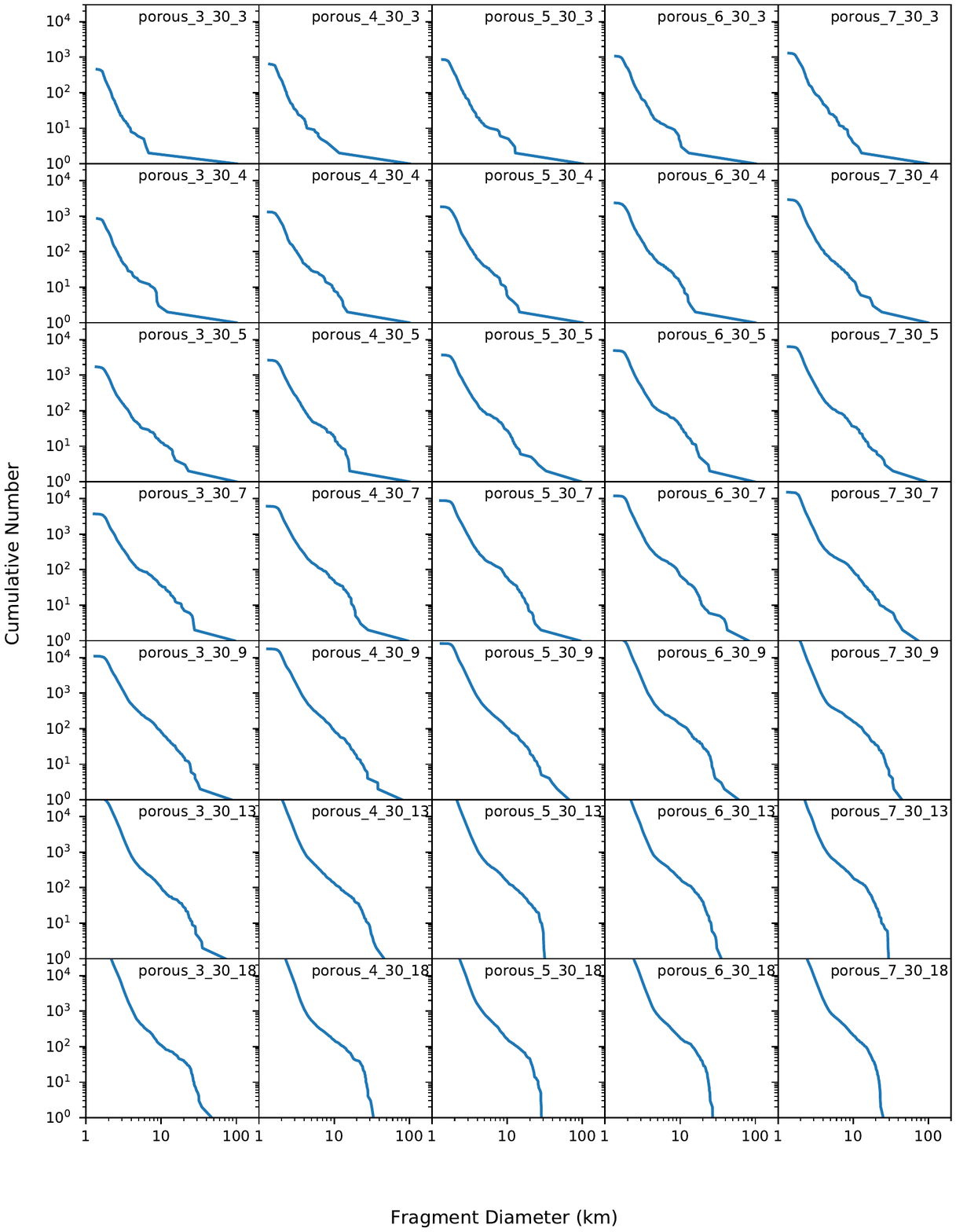}
\caption{Cumulative size distributions for impact angle $\theta$ = 30$^{\circ}$. For each run, the first number of the legend is the impact velocity in km/s, the second the impact angle and the third the approximate projectile radius in km.}
\label{fig:sfd30}
\end{center}
\end{figure}

\begin{figure}
\begin{center}
\includegraphics[width=14cm]{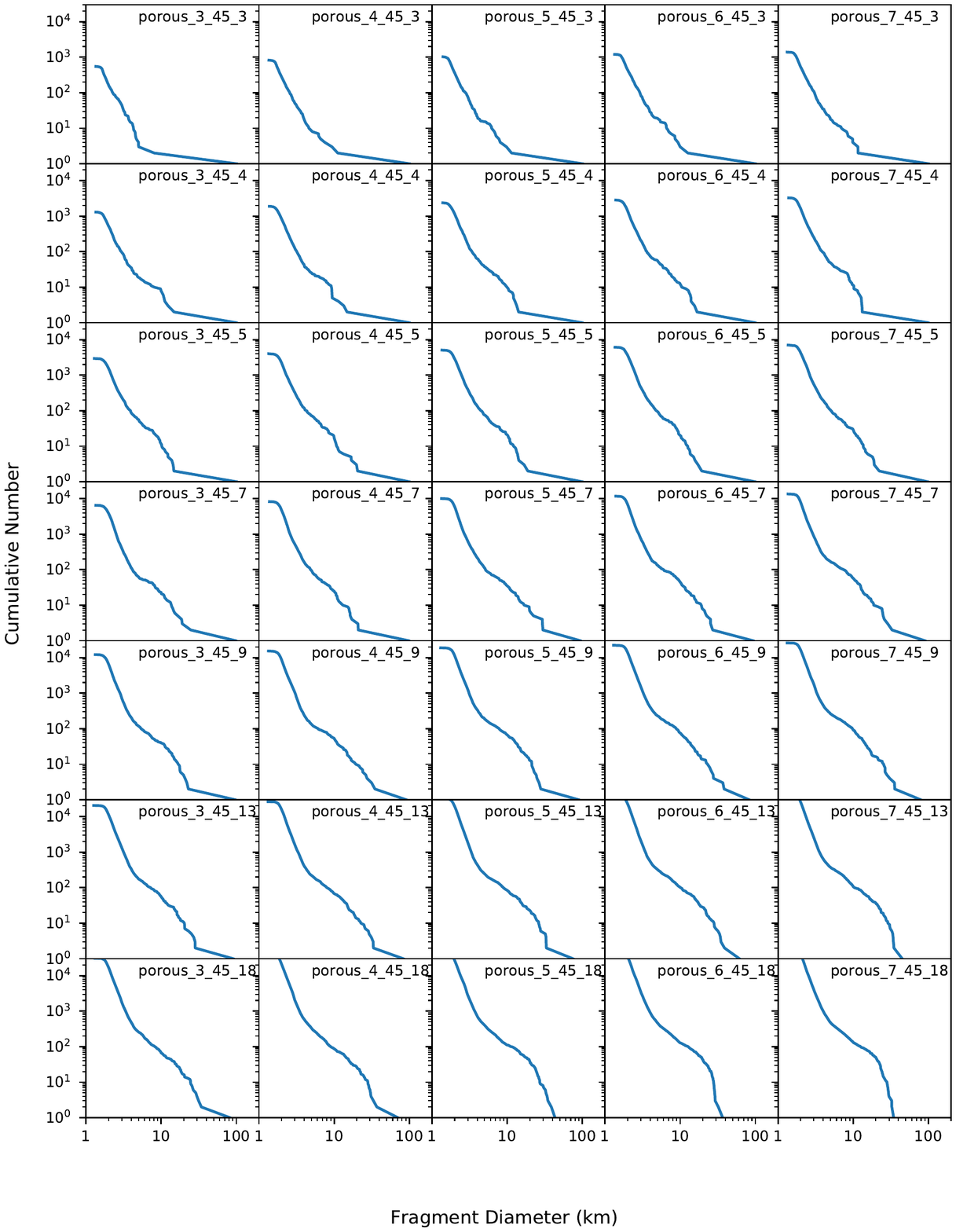}
\caption{Cumulative size distributions for impact angle $\theta$ = 45$^{\circ}$. For each run, the first number of the legend is the impact velocity in km/s, the second the impact angle and the third the approximate projectile radius in km.}
\label{fig:sfd45}
\end{center}
\end{figure}

\begin{figure}
\begin{center}
\includegraphics[width=14cm]{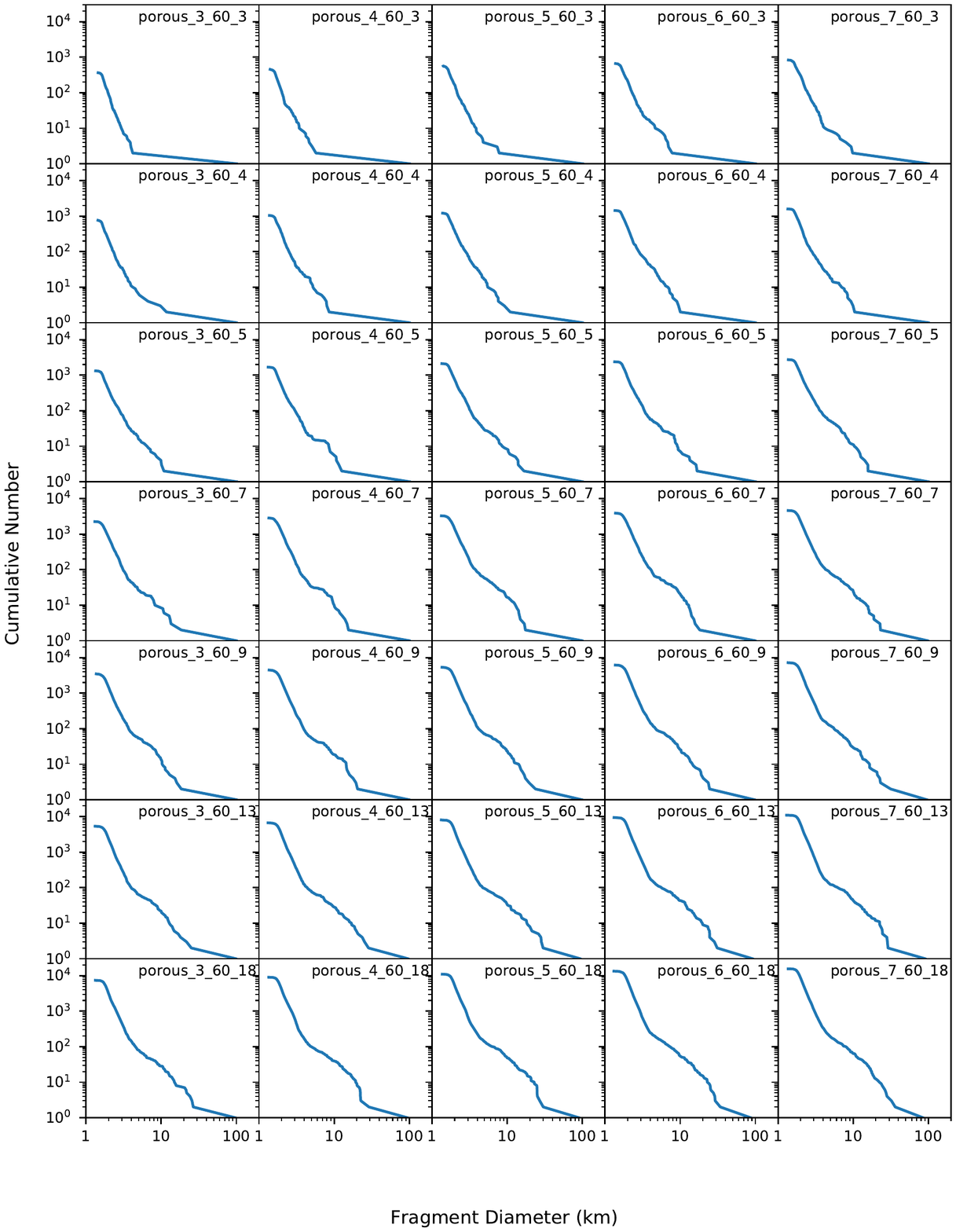}
\caption{Cumulative size distributions for impact angle $\theta$ = 60$^{\circ}$. For each run, the first number of the legend is the impact velocity in km/s, the second the impact angle and the third the approximate projectile radius in km.}
\label{fig:sfd60}
\end{center}
\end{figure}

\begin{figure}
\begin{center}
\includegraphics[width=14cm]{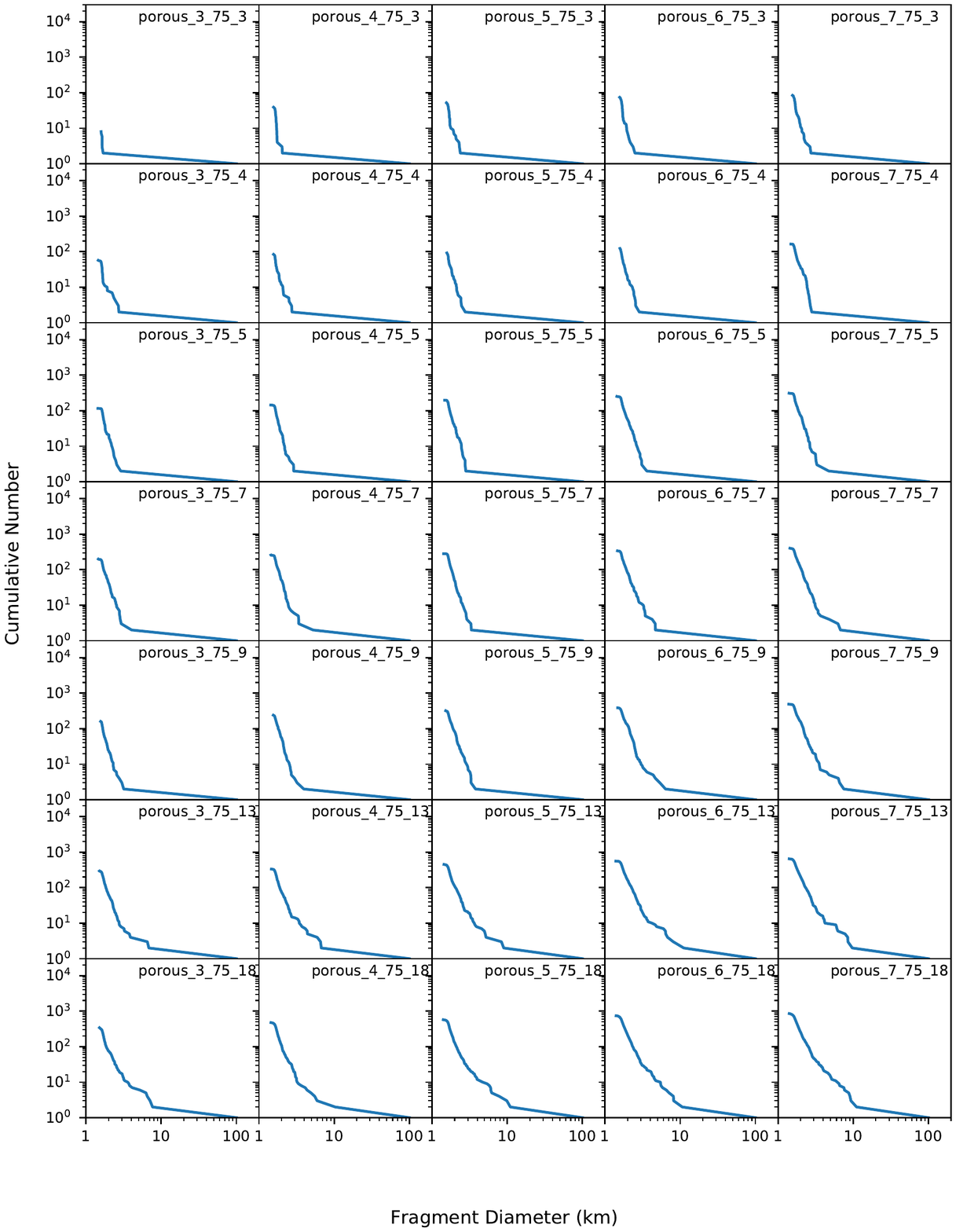}
\caption{Cumulative size distributions for impact angle $\theta$ = 75$^{\circ}$. For each run, the first number of the legend is the impact velocity in km/s, the second the impact angle and the third the approximate projectile radius in km.}
\label{fig:sfd75}
\end{center}
\end{figure}

The SFDs resulting from our simulations using porous targets are displayed in Figures \ref{fig:sfd15}-\ref{fig:sfd75}. As in previous studies, a wide range of morphologies is observed, depending on specific impact energy and impact angle. Low-energy and/or highly oblique impacts lead to cratering-type SFDs, while high-energy and/or close-to-head-on impacts lead to catastrophic or super-catastrophic disruptions (the corresponding size of largest remnant in each case is given in Figure  \ref{fig:mlrq}).

\subsubsection{Effect of parent body size}
\begin{figure}
\begin{center}
\includegraphics[width=14cm]{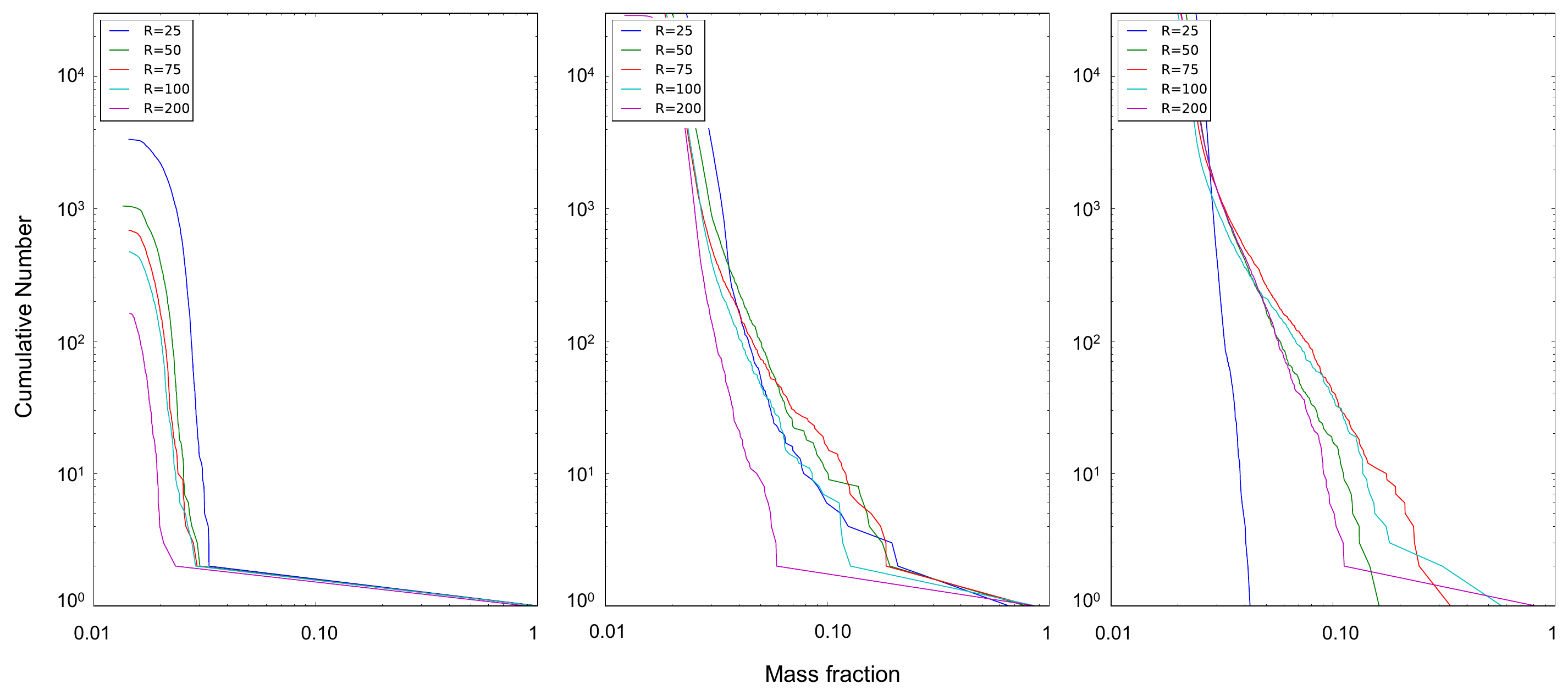}
\caption{Cumulative SFDs for different target sizes, normalized by target mass. Left: cratering regime; impact conditions correspond to case  "3$\_$45$\_$3". Middle: disruption regime,  case "3$\_$45$\_$18". Right: super-catastrophic regime, case  "7$\_$45$\_$18" (see Figure \ref{fig:sfd45} and main text for the different cases). In each regime, the same specific energies and mass ratios $M_p/M_t$ are used for the different target sizes.}
\label{fig:sfdcombined}
\end{center}
\end{figure}

Recent studies \citep{Benavidez:2018bd,Sevecek:2017sb} have investigated the dependence of the collision outcomes (such as the SFD) for given specific impact energies on the parent body size. In these studies, the results using two different targets sizes (either 100 km and 400 km; or 10 km and 100 km) were compared. Here, we systematically investigate the SFDs for 3 different impact regimes (cratering, disruption, super-catastrophic disruption) using a range of target radii (25, 50, 75, 100, 200 km). We consider the initial conditions of the cases "3$\_$45$\_$3", "3$\_$45$\_$18" and "7$\_$45$\_$18" (see Figure \ref{fig:sfd45}; the first number is the impact velocity in km/s, the second the impact angle and the third the approximate projectile radius in km) and adjust the projectile sizes to obtain the same specific impact energy for the various target sizes (i.e., the same mass ratio $M_p$/$M_t$ is used). Figure \ref{fig:sfdcombined}  displays the resulting size distributions. As expected, there is a clear dependence of the SFDs on the target size because of the change of $Q^*_D$ caused by the varying gravity potential \citep[e.g.][]{Jutzi:2015gb}.  Interestingly, the differences are more pronounced in the cratering and super-catastrophic regimes. Overall, these results confirm the findings by \citet{Benavidez:2018bd} and \citet{Sevecek:2017sb}, and strongly affirm that the 'linear scaling' of SFDs to different target sizes \citep{Durda:2007db} can only be applied over a limited size range.

\subsection{Velocity distributions}
\begin{figure}
\begin{center}
\includegraphics[width=14cm]{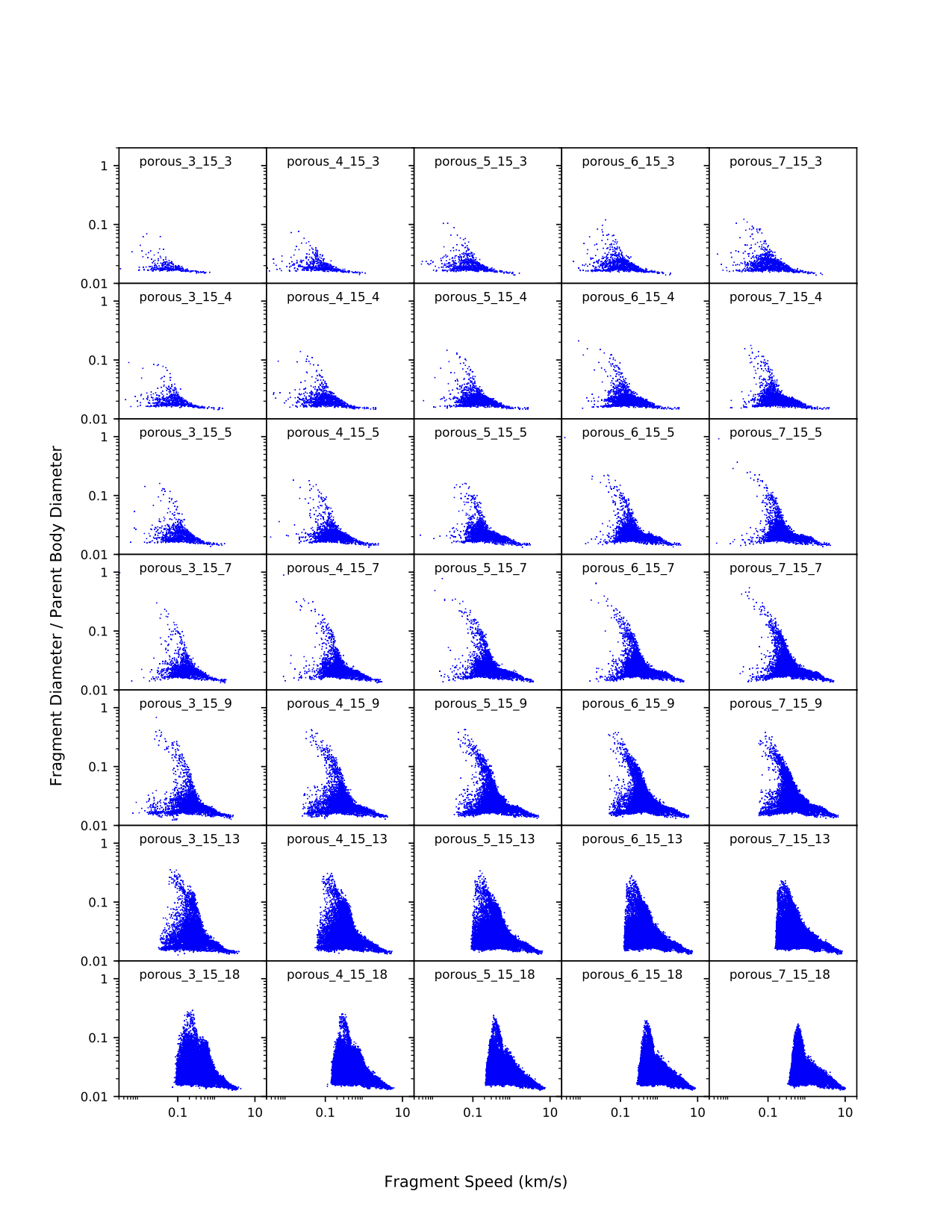}
\caption{Velocity distributions for impact angle $\theta$ = 15$^{\circ}$. For each run, the first number of the legend is the impact velocity in km/s, the second the impact angle and the third the approximate projectile radius in km.}
\label{fig:vel15}
\end{center}
\end{figure}

\begin{figure}
\begin{center}
\includegraphics[width=14cm]{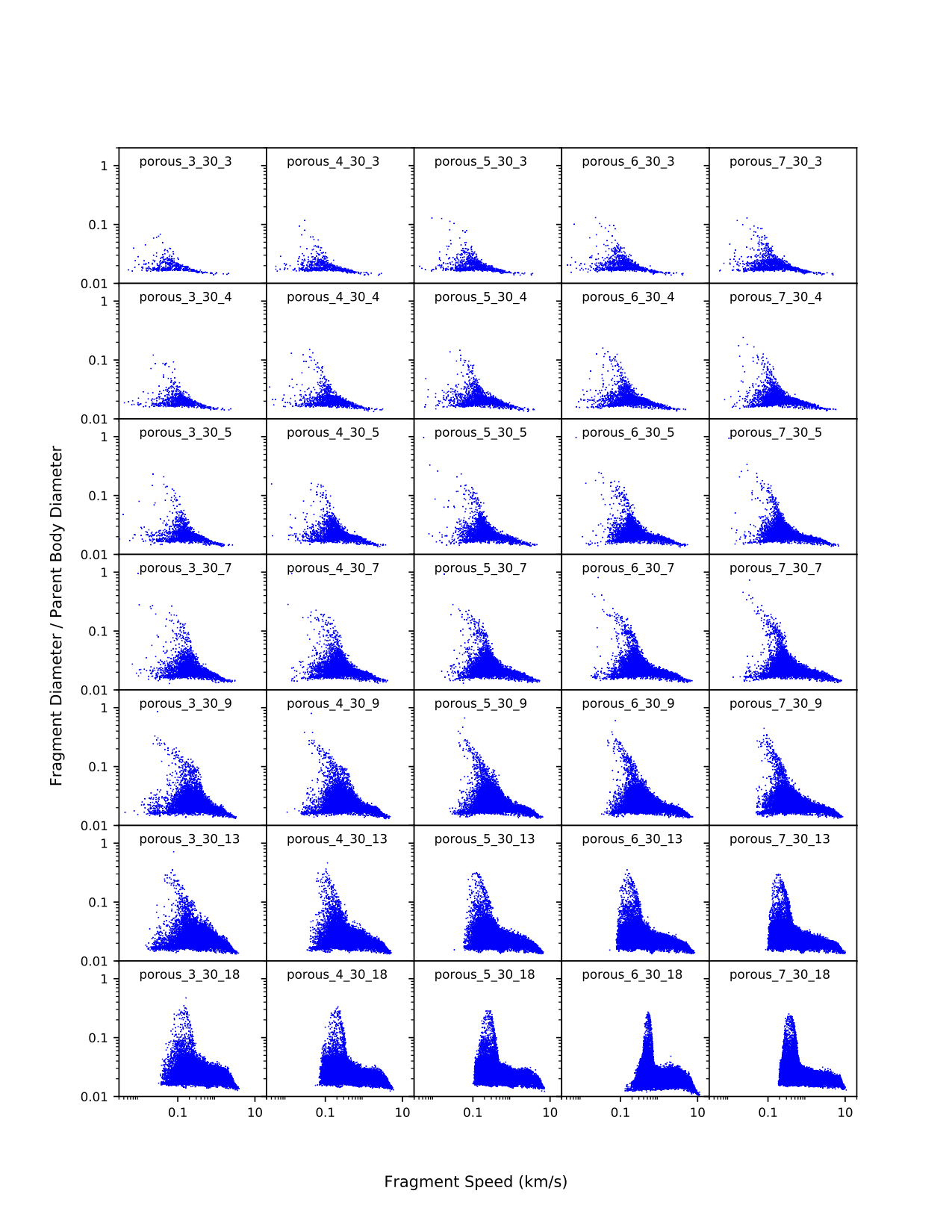}
\caption{Velocity distributions for impact angle $\theta$ = 30$^{\circ}$. For each run, the first number of the legend is the impact velocity in km/s, the second the impact angle and the third the approximate projectile radius in km.}
\label{fig:vel30}
\end{center}
\end{figure}

\begin{figure}
\begin{center}
\includegraphics[width=14cm]{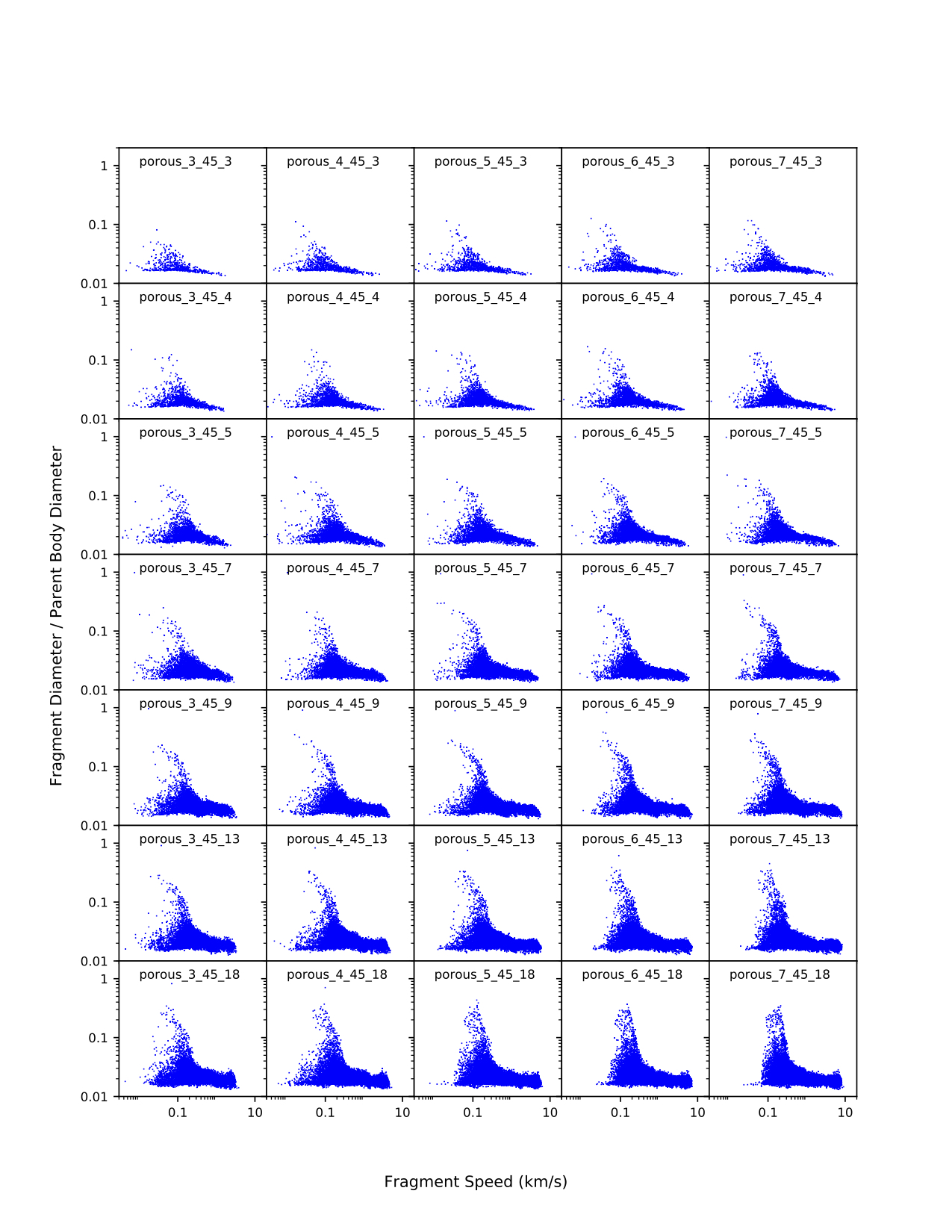}
\caption{Velocity distributions for impact angle $\theta$ = 45$^{\circ}$. For each run, the first number of the legend is the impact velocity in km/s, the second the impact angle and the third the approximate projectile radius in km.}
\label{fig:vel45}
\end{center}
\end{figure}

\begin{figure}
\begin{center}
\includegraphics[width=14cm]{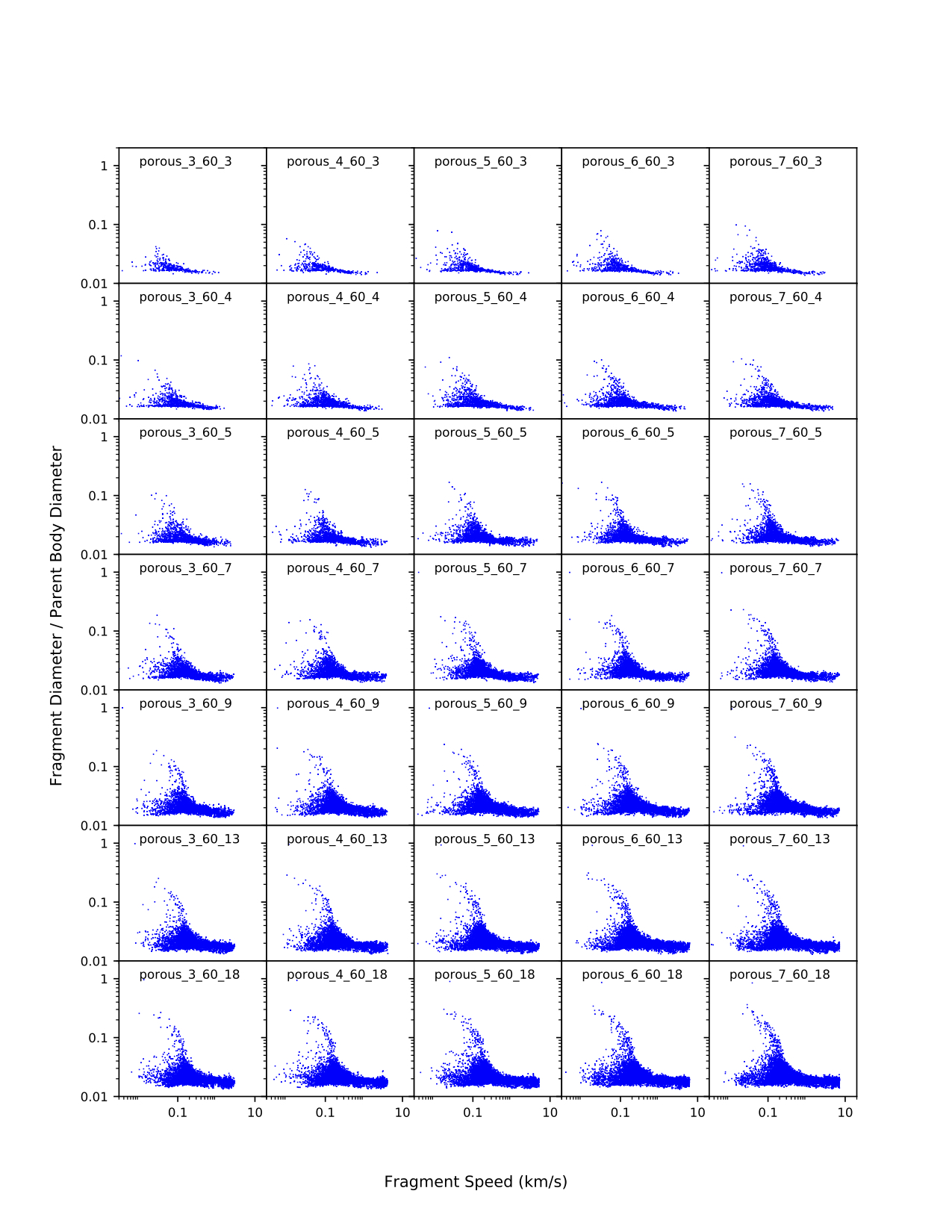}
\caption{Velocity distributions for impact angle $\theta$ = 60$^{\circ}$. For each run, the first number of the legend is the impact velocity in km/s, the second the impact angle and the third the approximate projectile radius in km.}
\label{fig:vel60}
\end{center}
\end{figure}

\begin{figure}
\begin{center}
\includegraphics[width=14cm]{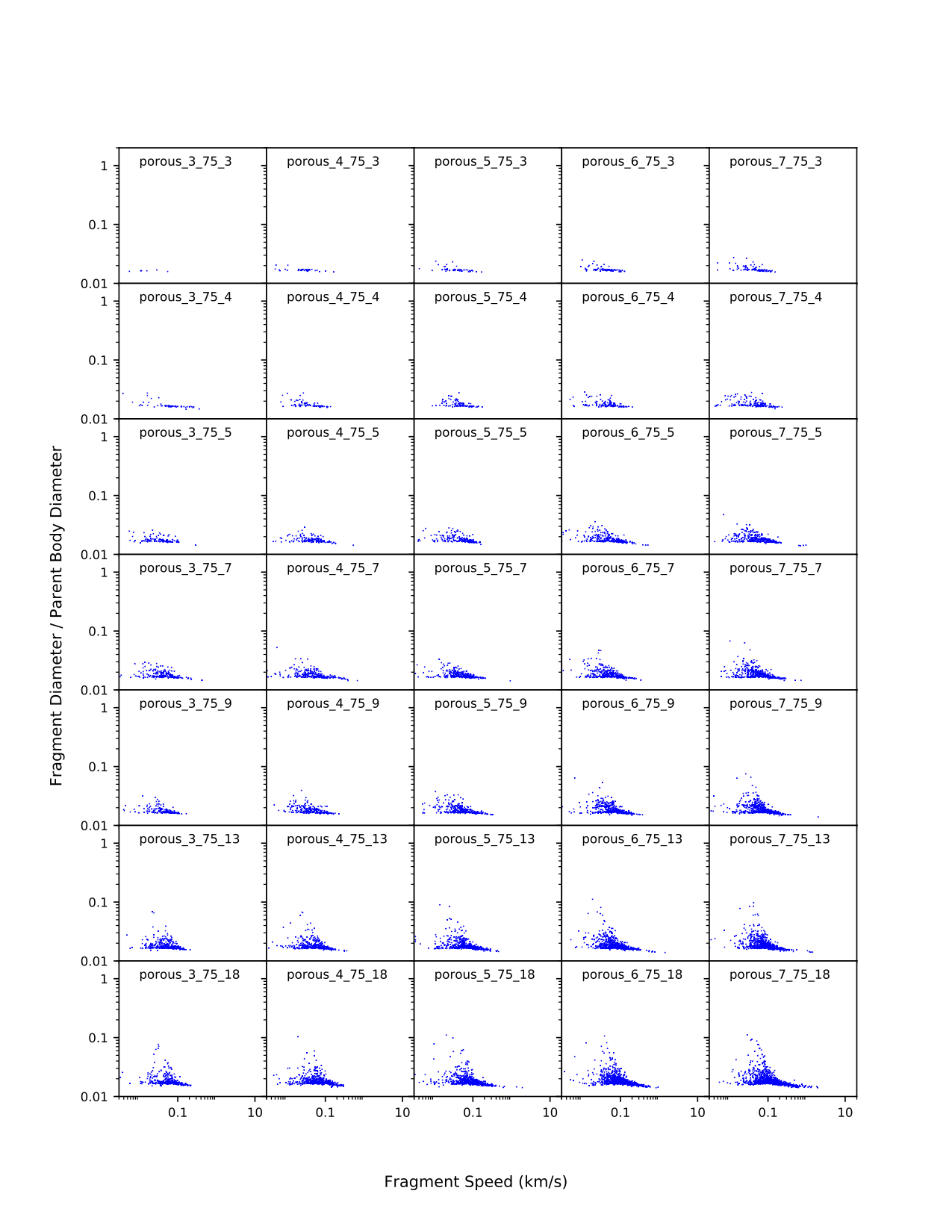}
\caption{Velocity distributions for impact angle $\theta$ = 75$^{\circ}$. For each run, the first number of the legend is the impact velocity in km/s, the second the impact angle and the third the approximate projectile radius in km.}
\label{fig:vel75}
\end{center}
\end{figure}

In addition to the SFDs we also compute the ejection velocity distributions (with respect to the target's center of mass). In Figures  \ref{fig:vel15} - \ref{fig:vel75}, the normalized fragment diameters are shown as a function of ejection velocity. Generally, there is a large range of ejection speeds, but only the small fragments reach high velocities. Interestingly, in some cases the highest ejection velocities are even larger than the impact velocity for a small fraction of the fragments. This may be related to the 'jetting effect', which can accelerate material to speeds larger than the impact velocity \citep[e.g.][]{Johnson:2014jb}. However, we note that some of these high velocity ejecta may be vaporized during ejection. Specific high-resolution simulations with more-sophisticated EOS models will be required to investigate this effect in more detail.

\section{Conclusions and outlook}\label{s:conclusions}
We have investigated the size-frequency distributions (SFDs) resulting from the disruption of 100 km-diameter targets consisting of porous material. Overall, the porous targets investigated here show a behavior more similar to the non-porous monolithic targets \citep{Durda:2007db} than the rubble-pile ones \citep{Benavidez:2012ez}, as they have a significantly higher impact strength ($Q^*_D$) than the latter \citep[see also][]{Jutzi:2015gb}. Various effects are important at these scales, such as self-gravity, material strength, friction and porosity, and partly compensate each other (e.g. higher material strength and densities (i.e. gravity) of non-porous objects vs. energy dissipation by pore-crushing of porous objects).  

Our results confirm that the SFDs resulting from a parent-body disruption are strongly dependent on the size scale (for a given specific impact energy), as shown in recent studies \citep{Benavidez:2018bd,Sevecek:2017sb}, and that the linear scaling approach is only valid over a limited size range. This emphasizes the need for additional studies, exploring a much larger range of the parameter space. 

The calculations and results presented here serve as a basis for a number of subsequent studies, comparing the SFDs to observed families (manuscript in prep.), as well as investigating the shapes of the largest remnants (Walsh et al., in prep.) and smaller ones (Barnouin et al., in prep.). They are also used  for the development of general scaling laws for small-body disruptions in the strength and gravity regime (Jutzi et al., in prep.). 

\section*{Acknowledgments}
M.J. acknowledges support from the Swiss National Centre of Competence in Research PlanetS. P.M. acknowledges support from the French space agency CNES.  D.C.R. acknowledges support from NASA grant NNX15AH90G awarded by the Solar System Workings program.

\bibliography{bibdata.bib}



\newpage

\end{document}